\documentclass[pre,aps,twocolumn]{revtex4}

\usepackage{amsmath,amssymb}
\usepackage{epsfig}

\newlength{\anchot}
\newlength{\anchof}
\newlength{\altof}
\newcommand{\fig}[1]{\parbox{\anchot}{\epsfig{file=fig/#1.eps,height=\altof}}}

\begin{document}

\title{Derivation of a Non-Local Interfacial Hamiltonian for Short-Ranged Wetting II:
General Diagrammatic Structure}

\author{A.\ O.\ Parry$^1$, C.\ Rasc\'{o}n$^2$, N.\ R.\ Bernardino $^1$, J.\ M.\ Romero-Enrique$^{3}$\\\
\vspace*{.75cm}}
\affiliation{$^1$ Department of Mathematics, Imperial College London,\\
London SW7 2BZ, United Kingdom \vspace*{.25cm} \\
$^2$ Grupo Interdisciplinar de Sistemas Complejos (GISC)\\
Dpto de Matem\'{a}ticas, Universidad Carlos III de Madrid\\ \vspace*{.15cm}
28911 Legan\'{e}s (Madrid), Spain\\
$^3$ Dpto de F\'{\i}sica At\'{o}mica, Molecular y Nuclear\\
Universidad de Sevilla, Ap.\ Correos 1065\\ \vspace*{.15cm}
41080 Seville, Spain}

\begin{abstract}
In our first paper, we showed how a non-local effective Hamiltionian for short-ranged wetting may be derived from an underlying Landau-Ginzburg-Wilson model. Here, we combine the Green's function method with standard perturbation theory to determine the general diagrammatic form of the binding potential functional beyond the double-parabola approximation for the Landau-Ginzburg-Wilson bulk potential. The main influence of cubic and quartic interactions is simply to alter the coefficients of the double parabola-like zig-zag diagrams and also to introduce curvature and tube-interaction corrections (also represented diagrammatically), which are of minor importance. Non-locality generates effective long-ranged many-body interfacial interactions due to the reflection of tube-like fluctuations from the wall. Alternative wall boundary conditions (with a surface field and enhancement) and the diagrammatic description of tricritical wetting are also discussed.

\end{abstract}


\maketitle

\section{Introduction}

In \cite{NoLo}, we showed how a non-local interfacial Hamiltonian for short-ranged wetting \cite{Review} may be derived from a Landau-Ginzburg-Wilson model using a diagrammatic formalism based on Green's functions \cite{Greens}. While the definition of the interfacial model is the same as that forwarded by other authors \cite{FJ1,FJ2,FJP}, its evaluation is non-perturbative in the interfacial gradient and reveals important non-local features. This has a number of advantages \cite{PRL,Jim1} over previous, local approximations, and appears to overcome a series of problems associated with short-ranged wetting \cite{FJ1,FJ2,PRL,Jim1,FJP,BHL,Binder,Covariance,Rejmer,Spheres,Jim2,ParryEvans}. The interaction of the interface and wall is described by a binding potential functional $W$ which has an elegant diagrammatic expansion
\begin{eqnarray}
W=a_1\fig{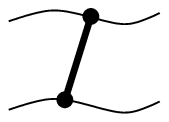}+b_1\fig{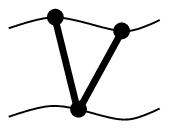}\,+\;\cdots
\label{1}
\end{eqnarray}
and an appealing physical interpretation as tube-like fluctuations \cite{Tubes} which zig-zag between the surfaces.

In this paper, we demonstrate the robustness of the diagrammatic expansion by extending the derivation beyond the double-parabola (DP) approximation for the bulk thermodynamic potential appearing in the Landau-Ginzburg-Wilson (LGW) Hamiltonian. We establish the general form of the asymptotic expansion of $W$, which now includes diagrams representing curvature corrections and tube-tube interactions. However, these are of negligible importance and the above diagrammatic expression remains valid albeit with numerical changes to the values of the coefficients $a_1$ and $b_1$, which can be determined exactly.

Our article is arranged as follows: After recapping briefly the central results of \cite{NoLo}, sections \ref{s3}-\ref{s6} describe the detailed derivation of $W$ beyond DP using the same diagrammatic formalism. The proof is rather technical and, to simplify things, we continue using fixed boundary conditions at the wall until the final section. Sections \ref{s7}-\ref{s9} are a lengthy discussion of the interpretation of the model and the evaluation of the diagrams for wetting transitions at planar walls which extends the analysis given in \cite{NoLo}. In particular, non-locality is shown to induce weak but long-ranged two-body interactions describing the repulsion of the interface from the wall. Alternative forms of boundary conditions, including coupling to an external surface field and enhancement \cite{NF}, and also diagrams describing tricritical wetting are also discussed.

\begin{figure}[h]
\vspace*{.35cm}\hspace*{0.2cm}\epsfig{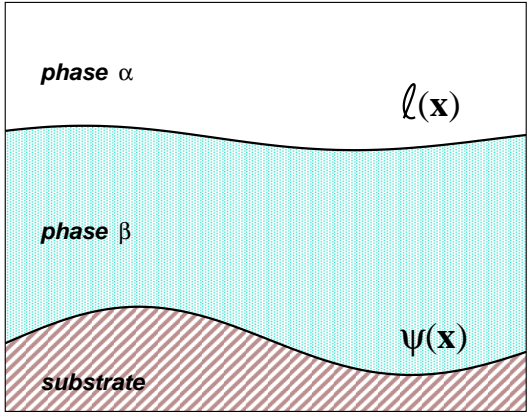}
\vspace*{.15cm}\caption{Schematic diagram showing a wetting layer of phase $\beta$ at a non-planar wall-$\alpha$ interface. Here, $\ell({\bf x})$ and $\psi({\bf x})$ are the interfacial collective coordinate and wall height, respectively.}
\end{figure}

\section{Wetting Diagrams within the Double-Parabola Approximation}
\label{s2}

We begin with some general considerations. Imagine a system bounded by a wall described by a height function $\psi({\bf{x}})$ measured above some reference plane with parallel vector displacement ${\bf{x}}=(x,y)$ (see Fig.1). The wall is in contact with a fluid phase $\alpha$ but preferentially adsorbs a fluid phase $\beta$ which forms a thin
film intervening between the bulk $\alpha$ phase and the wall. Throughout our paper, we will restrict our attention to the case of bulk two-phase coexistence, although it is easy to extend the analysis to non-zero values of the bulk ordering field. For systems with short-ranged forces, a convenient microscopic starting pointing for studying this is the continuum LGW Hamiltonian
\begin{equation}
H[m]=\int\!\!d{\bf r}\;\left\{\,\frac{1}{2}(\nabla m)^2+\Delta\phi(m)\,\right\}
\label{micro}
\end{equation}
based on a magnetization like order-parameter $m({\bf{r}})$. A bulk potential $\phi(m)$ describes the coexistence of the phases $\alpha$ and $\beta$ which, on imposing Ising symmetry, we identify with the spontaneous magnetizations
$-m_0$ and $+m_0$ respectively. The relative potential $\Delta\phi(m)=\phi(m)-\phi(m_0)$ is introduced to remove from the total free-energy a contribution proportional to the volume.

From (\ref{micro}), we wish to derive the form of an interfacial Hamiltonian pertinent to an "$m^4$" bulk potential
\begin{equation}
\phi(m)=-\frac{r}{2}\,m^2+\frac{u}{4}\,m^4
\end{equation}
below the bulk critical temperature. Bulk-like fluctuations are treated in mean-field fashion so the latter condition implies $r>0$. Written in terms of the mean-field spontaneous magnetization $m_0=\sqrt{r/u}$, and inverse bulk correlation length $\kappa=\sqrt{2r}$, the relative potential is
\begin{equation}
\Delta \phi(m)=\frac{\kappa^2}{\,8m_0^2\,}\,\left(m^2-m_0^2\right)^2
\label{m4}
\end{equation}
For later purposes, it is convenient to re-express this as
\begin{equation}
\Delta\phi(m)=\frac{\,\kappa^2}{2}\,\delta m ^2\,\left\{\,1+\frac{\delta m}{m_0}+
\frac{1}{4}\,\frac{\delta m^2}{m_0^2}\right\}
\label{m4expan}
\end{equation}
where we have defined $\delta m=|m|-m_0$. A much easier starting point for analysis is the DP approximation \cite{DP}
\begin{equation}
\Delta\phi^{(0)}(m)=\frac{\kappa^2}{2}\,(|m|-m_0)^2
\label{DP}
\end{equation}
in which one neglects the higher-order cubic and quartic terms \cite{FJ1}. The superscript indicates that the LGW model with a DP potential is the starting point for our perturbation theory. We define the reference Hamiltonian
\begin{equation}
H^{(0)}[m]=\int\!\!d{\bf{r}}\;\left\{\,\frac{1}{2}\,(\nabla m)^2+\Delta\phi^{(0)}(m)\right\}
\label{H0}
\end{equation}
Finally, we adopt the simplest choice of boundary condition corresponding to fixed surface magnetization. Thus, if $\,{\bf{r}}_{\psi}=({\bf{x}},\psi({\bf{x}}))$ denotes an arbitrary point on the wall, we require
\begin{equation}
m({\bf{r}}_{\psi})=m_1,
\end{equation}
for a fixed value of $m_1$. Without loss of generality, we set $m_1>0$ so that the wall preferentially adsorbs a thin layer of net positive ($\beta$-like) magnetization. For both the DP and the full "$m^4$" models, the mean-field wetting phase boundary, as defined for a planar wall, corresponds to $m_1=m_0$. That is, for $m_1<m_0$ the planar wall-$\beta$ interface is partially wet (with finite contact angle $\theta$) and the $\alpha|\beta$ interface unbinds continuously as $m_1\to m_0^-$. The wall is completely wet by the $\beta$ phase for $m_1>m_0$ corresponding to $\theta=0$. Note that, with these boundary conditions, first-order and tricritical wetting transitions do not arise. It is convienient to introduce a dimensionless temperature-like scaling field
\begin{equation}
t=\frac{\,m_1-m_0\,}{m_0}
\end{equation}
which measures the deviation from the three-dimensional critical wetting phase boundary.

\begin{figure}[t]
\vspace*{.35cm}\hspace*{0.2cm}\epsfig{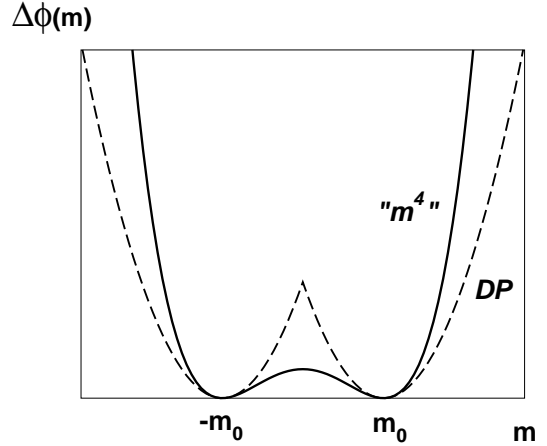}
\vspace*{.15cm}\caption{The relative potential $\Delta\phi(m)$ in the "$m^4$" theory (\ref{m4}) and DP approximation (\ref{DP}) at bulk coexistence.}
\end{figure}

Following Fisher and Jin \cite{FJ1,FJ2} we define the interfacial co-ordinate $\ell({\bf{x}})$ using a crossing-criterion and consider constrained profiles which have a surface of iso-magnetization $m^X=0$ at some prescribed interfacial configuration:
\begin{equation}
m({\bf{r}}_{\ell})=0
\label{crossing}
\end{equation}
for all points ${\bf{r}}_{\ell}=({\bf{x}},\ell({\bf{x}}))$. Starting from a suitable microscopic model $H[m]$, the interfacial Hamiltonian $H[\ell,\psi]$ is defined via a partial trace over this class of constrained profiles. A saddle point approximation leads to the Fisher-Jin identification
\begin{equation}
H[\ell,\psi]=H[m_{\Xi}({\bf{r}})]-F_{w\beta}[\psi]
\label{FJident}
\end{equation}
where $F_{w\beta}[\psi]$ is the excess-free energy of the wall-$\beta$ interface. Here $m_{\Xi}({\bf{r}})$ is the constrained profile that minimises the LGW Hamiltonian with the appropriate boundary conditions at the interface, wall and in the bulk. That is, the constrained profile satisfies the variational equation
\begin{equation}
\frac{\delta H[m]}{\delta m}\Bigg|_{m_{\Xi}({\bf{r}})}=0
\label{var}
\end{equation}
We now focus on the properties of the the DP model (\ref{H0}) summarising the main results of our previous article. The variational equation (\ref{var}) leads to the Helmholz equation
\begin{equation}
\nabla^2 \delta m_{\Xi}^{(0)}=\kappa^2\delta m_{\Xi}^{(0)}
\label{Helm}
\end{equation}
where $\delta m_{\Xi}\equiv |m_{\Xi}|-m_0$. Again, the superscript (not used in \cite{NoLo}) serves to indicate that the DP potential is the zeroth-order term in a perturbative expansion. The linearity of these equations simplifies considerably the derivation of the non-local model. Making use of the divergence theorem,
\begin{eqnarray}
\label{div}
H^{(0)}[m_{\Xi}^{(0)}]=-\frac{\delta m_1}{2}\int_{\psi}\!\!d{\bf{s}}_{\psi}\;\,\nabla
m_{\Xi}^{(0)}\cdot{\bf{n}}_{\psi}
\\ \nonumber
-\frac{ m_0}{2}\int_{\ell^-}\!\!\!\!d{\bf{s}}_{\ell}\;\,\nabla
m_{\Xi}^{(0)}\cdot{\bf{n}}_{\ell}-
\frac{m_0}{2}\int_{\ell^+}\!\!\!d{\bf{s}}_{\ell}\;\,\nabla
m_{\Xi}^{(0)}\cdot{\bf{n}}_{\ell}
\end{eqnarray}
contains only surface terms. Here, $\delta m_1 = m_1 - m_0$ while $\bf{n}_{\psi}$ and $\bf{n}_{\ell}$ denote
the (local) unit normals at the wall and interface respectively, pointing into the bulk. Similarly, the infinitesimals $d{\bf{s}}_{\psi}$ and $d{\bf{s}}_{\ell}$ represent local area elements at the wall and interface. The solutions to the Helmholtz equations in the bulk ($m_\Xi < 0$) and wetting ($m_\Xi > 0$) regions are written in terms of the Green's function
\begin{equation}
K({\bf{r}}_1,{\bf{r}}_2)=\frac{\kappa}{\,2\pi|{\bf r}_1-{\bf r}_2|\,}\;\,
e^{-\kappa\,|{\bf{r}}_1-{\bf{r}}_2|}
\label{K}
\end{equation}
which satisfies the Ornstein-Zernike like equation
\begin{equation}
(-\nabla^2_{{\bf{r}}_1}+\kappa^2)\;K({\bf{r}}_1,{\bf{r}}_2)
\,=\,2\,\kappa\,\delta({\bf{r}}_1-{\bf{r}}_2)
\label{OZ}
\end{equation}
and decays to zero as $\vert {\bf r}_1 -{\bf r}_2 \vert \to \infty $. We represent this Green's function diagrammatically by a straight thick line with the open circles denoting the end points
\begin{equation}
K({\bf{r}}_1,{\bf{r}}_2)=\fig{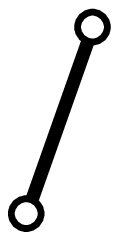}
\end{equation}
Using the Green's function, we identify the constrained magnetization $m_{\Xi}^{(0)}$ in the bulk region
\begin{equation}
m_{\Xi}^{(0)}=-m_0+m_0\,\fig{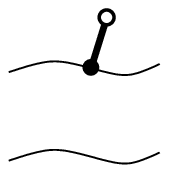}
\end{equation}
and, within the wetting layer, via the expansion
\begin{eqnarray}
\delta m_{\Xi}^{(0)}=-m_0\left(\fig{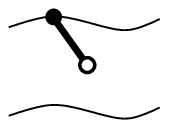}-\fig{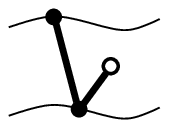}+\fig{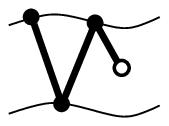}-\;\cdots\right)
\nonumber\\
+\delta m_1\left(\fig{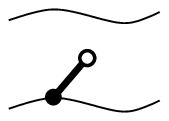}-\fig{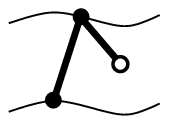}+\fig{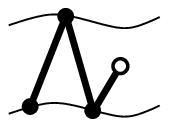}-\;\cdots\right)\qquad\quad
\label{DPM}
\end{eqnarray} 
In this diagrammatic notation, the wavy lines represent the constrained interfacial configuration (top) and wall (bottom), while a black dot on a surface means one must integrate over all points on that surface with the appropriate infinitesmal area element. These expressions are exact solutions to the Helmholtz equations, and satisfy the boundary conditions at the interface and wall to exponentially accurate order in the radii of curvature.

After substituting into (\ref{div}) and making use of the method of images, we arrive at the desired result
\begin{equation}
H^{(0)}[\ell,\psi]=H^{(0)}_{\alpha\beta}[\ell]+W^{(0)}[\ell,\psi]
\end{equation}
where
\begin{equation}
H^{(0)}_{\alpha\beta}[\ell]=\Sigma^{(0)}_{\alpha\beta}\;A_{\alpha\beta}
\label{DPFree}
\end{equation}
is the interfacial Hamiltonian of the free $\alpha|\beta$ interface, $A_{\alpha\beta}$ is the interfacial area and $\Sigma^{(0)}_{\alpha\beta} = \kappa m_0^2$ is the DP result for the interfacial tension. The binding potential functional is given by
\begin{equation}
W^{(0)}[\ell,\psi]=\sum_{n=1}^{\infty}\,\left\{\,
a_1^{(0)}\,\Omega_n^n+b_1^{(0)}\,\Omega_n^{n+1}+b_2^{(0)}\,\Omega_{n+1}^n\right\}
\label{DPW}
\end{equation}
with geometry independent coefficients
\begin{equation}
\frac{a_1^{(0)}}{\kappa m_0^2}=2\,t,\qquad\frac{b_1^{(0)}}{\kappa m_0^2}=1,\qquad\frac{b_2^{(0)}}{\kappa m_0^2}=t^2
\label{DPBare}
\end{equation}
and $\Omega_\mu^\nu$ correspond to surface integrals over products of the kernel $K$. The diagrammatic and algebraic representations of the first three terms are
\begin{eqnarray}
\Omega_1^1[\ell,\psi]=\fig{001}
=\int\!\!\!\!\int d{\bf{s}}_{\psi}\,d{\bf{s}}_{\ell}\;K({\bf{r}}_{\psi},{\bf{r}}_{\ell})
\label{24}
\end{eqnarray}
\begin{eqnarray}
\Omega_1^2[\ell,\psi]=\fig{004}
=\int\!\!d{\bf{s}}_{\psi}\,\Bigg\{\!\int\!\!d{\bf{s}}_{\ell}\;K({\bf{r}}_{\psi},{\bf{r}}_{\ell})\Bigg\}^2
\label{25}
\end{eqnarray}
\begin{eqnarray}
\Omega_2^1[\ell,\psi]=\fig{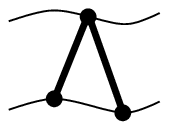}
=\int\!\!d{\bf{s}}_{\ell}\,\Bigg\{\!\int\!\!d{\bf{s}}_{\psi}\;K(\bf{r}_{\psi},{\bf r}_{\ell})\Bigg\}^2
\label{26}
\end{eqnarray}
although for certain configurations further simplification is possible (see later). All diagrams have a zig-zag structure, for example,
\begin{eqnarray}
\Omega_2^2[\ell,\psi]=\fig{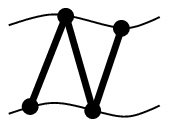},\qquad\Omega_2^3[\ell,\psi]=\fig{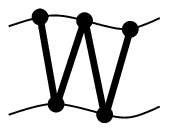}
\nonumber
\end{eqnarray}
Thus, up to "two tubes", the asymptotic expansion of $W$ is
\begin{eqnarray}
W^{(0)}=a^{(0)}_1\fig{001}+b^{(0)}_1\fig{004}+b^{(0)}_2\fig{005}+\cdots
\label{27}
\end{eqnarray}
with the higher-order terms resumming to give the hard-wall repulsion of the wall.

The coefficient of each diagram $\Omega_{\mu}^{\nu}$ has the dimensions of a surface tension and exhibits a power-law dependence on the scaling field $t$. A handy rule to remember is that the power of $t$ is the same as the number of black dots on the wall that are singly connected, {\sl i.e.\ }have only one kernel $K$ attached to it. This rule also applies to all diagrams containing only black dots which will be generated in the perturbation series described next.

One also generates an expression for the excess free-energy of the wall-$\beta$ interface
\begin{equation}
F_{w\beta}^{(0)}[\psi]= \Sigma_{w\beta}^{(0)}\,A_w+C_{w\beta}^{(0)}\int\!\!
d{\bf{s}}_{\psi}\,\left\{\frac{1}{R_1^{\psi}}+\frac{1}{R_2^{\psi}}\right\}
\label{DPFwbeta}
\end{equation}
which depends on the area $A_w$ and the mean-curvature of the wall. The latter are expressed in terms of the local principle radii of curvature $R_1^{\psi}$ and $R_2^{\psi}$. The tension and rigidity are given by $\Sigma_{w\beta}^{(0)}=\kappa\,m_0^2\,t^2/2$ and $C_{w\beta}^{(0)}= m_0^2\,t^2/4$ respectively.

Finally, for wetting at planar substrates ($\psi=0$) the non-local model recovers the known form of the approximate local interfacial Hamiltonian when $\nabla \ell\ll1$. We find
\begin{equation}
H^{(0)}[\ell]\approx\int\!\!d{\bf{x}}
\left\{\frac{\Sigma^{(0)}({\ell})}{2}\,(\nabla\ell)^2+W^{(0)}_\pi({\ell})\right\}+\Sigma^{(0)}_{\alpha\beta}\,A_w
\label{FJ}
\end{equation}
where $W_\pi(\ell)$ and $\Sigma({\ell})=\Sigma_{\alpha\beta}+\Delta\Sigma({\ell})$ are the binding potential function and effective position-dependent stiffness, respectively. Within the DP approximation,
\begin{equation}
W^{(0)}_\pi({\ell})=\;w_{10}^{(0)}\,e^{-\kappa\ell}+w_{20}^{(0)}\,e^{-2\kappa\ell}+\;\cdots
\end{equation}
and
\begin{equation}
\Delta\Sigma^{(0)}({\ell})=\,s_{10}^{(0)}\,e^{-\kappa\ell}+s_{21}^{(0)}\kappa\ell\,e^{-2\kappa\ell}+\;\cdots
\end{equation}
and are identical to the findings of Fisher and Jin \cite{FJ1,FJ2}, who derived the small gradient (local) limit (\ref{FJ}) directly. The coefficients appearing in these expressions are determined by the coefficients of the binding potential functional. Thus, we identify $w_{10}^{(0)}=s_{10}^{(0)}=a_1^{(0)}$ while $w_{20}^{(0)}=b_1^{(0)}+b_2^{(0)}$, and $s_{21}^{(0)}=-2b_1^{(0)}$. We now wish to see how the above results are altered when one goes beyond the DP approximation.

\section{Feynman-Hellman Theorem and Perturbation Theory}
\label{s3}

Let us suppose that our microscopic model $H[m]$ can be written 
\begin{equation}
H[m]=H^{(0)}[m]+\epsilon\, H^{(1)}[m]
\end{equation}
containing a dimensionless field $\epsilon$ which will later act as a small parameter.  The reference Hamiltonian is the DP model, while
\begin{equation}
H^{(1)}[m]=\int\!\!d{\bf{r}}\;\Delta \Phi^{(1)}(m)
\label{FH}
\end{equation}
accounts for cubic and quartic corrections obtained by writing
\begin{equation}
\Delta \phi(m)=\frac{\,\kappa^2\,\left(|m|-m_0\right)^2}{2}+\epsilon\,\Delta\Phi^{(1)}(m)
\label{phi0+phi1}
\end{equation}
with
\begin{equation}
\Delta\Phi^{(1)}(m)=\frac{\kappa^2}{2}\,\delta m ^2\,\left(\,\frac{\delta m}{m_0}+
\frac{1}{4}\,\frac{\delta m^2}{m_0^2}\right).
\end{equation}
Thus, the potential (\ref{phi0+phi1}) interpolates between the DP model ($\epsilon=0$) and the "$m^4$" model ($\epsilon=1$). Recall that the interfacial model is identified by evaluating $H[m]$ at the constrained magnetization $m_\Xi$. Taking the derivative of the constrained Hamiltonian
\begin{equation}
\frac{dH[m_{\Xi}]}{d\epsilon}=H^{(1)}[m_{\Xi}]+\int\!\!d{\bf{r}}\; \left.\frac{\delta H}{\delta m} \right|_{m_{\Xi}} \frac{dm_{\Xi}}{d \epsilon}
\end{equation}
which, by virtue of the variational condition (\ref{var}), leads to the familiar expression
\begin{equation}
\frac{dH[m_{\Xi}]}{d\epsilon}=\int\!\! d{\bf{r}}\;\Delta \Phi^{(1)}(m_{\Xi}),
\label{FH}
\end{equation}
similar to the well known Feynman-Hellman theorem in standard quantum mechanics. Note that the functional on the R.H.S depends on the full ($\epsilon$ dependent) constrained magnetization and is a convenient means of formulating a perturbation expansion
\begin{equation}
H[m_{\Xi}]=H^{(0)}[m_{\Xi}^{(0)}]+\epsilon\,\tilde H^{(1)}+\epsilon^2\, \tilde H^{(2)}+\;\cdots
\end{equation}
From this, it is straightforward to determine the corresponding expansion for the binding potential functional
\begin{equation}
W[\ell,\psi]=W^{(0)}[\ell,\psi]+\epsilon\, W^{(1)}[\ell,\psi]+\epsilon^2\, W^{(2)}[\ell,\psi]+\;\cdots
\end{equation}
where the leading order-term is the DP result (\ref{DPW}). In addition, we will also be able to compute expansions for the free interface $H_{\alpha\beta}[\ell]$ and the excess free-energy of the wall-$\beta$ interface $F_{w\beta}[\psi]$.

To determine the first-order and second-order perturbation functionals $\tilde H^{(1)}$ and $\tilde H^{(2)}$, we return to the Euler-Lagrange equation for the constrained profile
\begin {equation}
\nabla^2 \delta m_{\Xi}=\kappa^2 \delta m_{\Xi}+\epsilon\,\frac{\,\partial\Delta\Phi^{(1)}(m_{\Xi})}{\partial m}
\end{equation}
and seek a perturbative solution
\begin{equation}
\delta m_{\Xi}({\bf{r}}; \epsilon)=\delta m_{\Xi}^{(0)}({\bf{r}})+\epsilon\;\delta m_{\Xi}^{(1)}({\bf{r}})+\;\cdots
\end{equation}
By definition, the leading-order term is the DP result, which satisfies the Helmholtz equation (\ref{Helm}), while the first-order correction satisfies the inhomogeneous PDE
\begin{equation}
\nabla^2 \delta m_{\Xi}^{(1)}=\kappa^2\, \delta m_{\Xi}^{(1)}+\frac{\partial\Delta
\Phi^{(1)}(m_{\Xi}^{(0)})}{\partial m}
\end{equation}
and vanishes at the interface, wall, and at infinity. Combining these, we obtain
\begin{equation}
\tilde H^{(1)}[\ell,\psi]=\int\!\!d{\bf{r}}\; \Delta\Phi^{(1)}(m_{\Xi}^{(0)})
\label{Pert1}
\end{equation}
and
\begin{equation}
\tilde H^{(2)}[\ell,\psi]=\frac{1}{2}\int\!\!d{\bf{r}}\;\delta m_{\Xi}^{(1)}\,\frac{\,\partial \Delta\Phi^{(1)}
(m_{\Xi}^{(0)})}{\partial m}
\label{Pert2}
\end{equation}
A simplifying feature of the first-order correction is that it only depends on the zeroth-order profile $m_{\Xi}^{(0)}$ as calculated within the DP approximation. We begin with such a calculation for some preliminary quantities.

\section{First-order Perturbation theory for the Free Hamiltonian}
\label{s4}

Consider a free but constrained configuration of the $\alpha-\beta$ interface. That is, the interface is infinitely far from any confining wall but the magnetization is constrained to be zero along a surface at height $\ell({\bf{x}})$. Bulk phases $\alpha$ and $\beta$ lie above and below the interface respectively. The interface partitions the system into two regions whose order-parameter fluctuations are shielded from each other, by virtue of the crossing-criterion. The zeroth-order DP expressions for the position-dependent magnetizations in these regions are
\begin{eqnarray}
m_{\Xi}^{(0)}({\bf{r}})=-m_0+m_0\fig{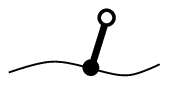}
\end{eqnarray}
and
\begin{eqnarray}
m_{\Xi}^{(0)}({\bf{r}})=m_0-m_0\fig{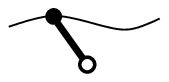}
\end{eqnarray}
above and below the interface respectively. The first-order result for the free interfacial Hamiltonian is
\begin{equation}
H_{\alpha\beta}[\ell]=H_{\alpha\beta}^{(0)}[\ell] +\epsilon\int\!\!d{\bf{r}}\; \Delta \Phi^{(1)} (m_{\Xi}^{(0)})+\;\cdots
\end{equation}
where the first term is simply the zeroth-order DP result $\,H_{\alpha\beta}^{(0)}[\ell]=\Sigma_{\alpha\beta}^{(0)}\, A_{\alpha\beta}$. Hence,
\begin{eqnarray}
H_{\alpha\beta}[\ell]=\Sigma_{\alpha\beta}^{(0)}\,A_{\alpha\beta}+
\hspace*{5.5cm}\nonumber\\
\epsilon\,\kappa\,m_0^2\, \left\{-\frac{1}{2}\fig{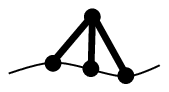}\!-\frac{1}{2}\fig{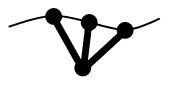}\!+\frac{1}{8}\fig{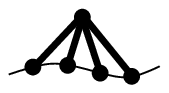}\!+\frac{1}{8}\fig{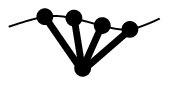}\!\!\right\}
\end{eqnarray}
where we have expressed the results diagrammatically. The single wavy line represents the free interface while the thick straight lines denote the Green's function $K$. The diagrams appearing in this formula are all of the same type and have $n=3,4$ (black) dots on the interface and one (black) dot either above or below the surface. They correspond
to multi-dimensional integrals. For example,
\begin{equation}
\fig{022}= \kappa\int_{V_+}\! d{\bf{r}}\; \left\{\int\!\!d{\bf{s}}_{\ell}\; K({\bf{r}}_{\ell},{\bf{r}})\,\right\}^4
\end{equation}
where, in general, the integrand contains $n$ Kernels $K$ connecting a point off the interface to $n$ different points on it. Black dots on the surface have the same intepretation as before - one must integrate over all points on the surface with the appropriate area element. A black point off the surface means that one must integrate over the appropriate semi-volume $V_+$(here above the interface) together with a multiplicative factor of $\kappa$. The latter is introduced so the diagram has the dimensions of area. Again, each Kernel may be thought of as representing a short tube-like fluctuation protruding from the surface, only a few bulk correlation lengths long (since the Kernel decays exponentially quickly). Such fluctuations can be thought of as giving the interface a "corona". As we shall show, these shift the DP expression for the surface tension and also introduce curvature corrections. To see this, consider first the case of a planar interface of (infinite) area $A_{\alpha\beta}$. By definition, the value of the Hamiltonian per unit area is equal to the surface tension, so we can identify
\begin{equation}
\Sigma_{\alpha\beta} (\epsilon)=\Sigma_{\alpha\beta}^{(0)}
+\frac{\,\kappa m_0^2\,\epsilon\,}{A_{\alpha\beta}}\,\left\{-\fig{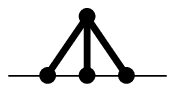}+\frac{1}{4}\fig{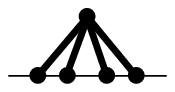}\right\}
\end{equation}
 The integrals are easily performed
\begin{equation}
\fig{024}=\frac{A_{\alpha\beta}}{3}\qquad
\fig{025}=\frac{A_{\alpha\beta}}{4}
\end{equation}
which implies the tension is shifted to
\begin{equation}
\Sigma_{\alpha\beta}(\epsilon)=\kappa\,m_0^2\,\left\{\,1+\epsilon\left(-\frac{1}{3}+\frac{1}{16}\right)+\cdots\,\right\}
\label{MFTensionfirst}
\end{equation}
where we have highlighted the different numerical contributions for the cubic and quartic perturbations. Setting $\epsilon=1$, we find $\Sigma_{\alpha\beta}\approx 0.72\,\kappa\,m_0^2$, which is in much better agreement with the mean-field expression $\Sigma_{\alpha\beta}=(2/3)\kappa m_0^2$ of the full "$m^4$" theory \cite{RW}. Thus, the dominant numerical correction to the DP expression for the surface tension arises from the cubic term in $\Delta\phi^{(1)}$ and is accurately accounted for by first-order pertubation theory. This point is further amplified by calculating exactly the mean-field surface tension $\Sigma_{\alpha\beta}(\epsilon)$ for the potential (\ref{phi0+phi1}):
\begin{eqnarray}
\frac{\Sigma_{\alpha\beta}(\epsilon)}{\kappa m_0^2}= \left(\frac{4}{3\epsilon}-2\right)\left(\sqrt{4-3\epsilon\,}-2\right)
\\ \nonumber
+\frac{4(1-\epsilon)}{\sqrt {\epsilon}}\, \ln\frac{2(1+\sqrt{\epsilon})}{\,\sqrt{4-3\epsilon}+\sqrt {\epsilon}\,}
\hspace*{1.5cm}
\end{eqnarray}
It is straightforward to check that this is consistent with the limiting cases at $\epsilon=1$ and
$\epsilon=0$ respectively, and also reproduces the perturbation expansion (\ref{MFTensionfirst}). While this function looks rather ominous, it is almost linear in character over the required domain.

In addition to correcting the value of the surface tension, the "corona" diagrams lead to curvature terms, which reveal the more general structure of the free Hamiltonian. To appreciate this, consider the case of an undulating interfacial profile. Provided the local principal radii of curvature $R_1^{\,\ell}({\bf{x}}), R_2^{\,\ell}({\bf{x}})$, are always large, one can expand the integrals to find
\begin{equation}
H_{\alpha\beta}[\ell]=\!\int\!\!d{\bf{s}}_{\ell}\, \left\{\Sigma_{\alpha\beta}
+\frac{\kappa_{\alpha\beta}}{2}\left(\frac{1}{R_1^{\ell}}+\frac{1}{R_2^{\ell}}\right)^2\!\!
+\frac{\bar\kappa_{\alpha\beta}}{\,R_1^{\ell}R_2^{\ell}\,}+\;\cdots\right\}
\end{equation}
where $\kappa_{\alpha\beta}\!=\!\epsilon m_0^2/64\kappa$, and $\bar\kappa_{\alpha\beta}\!=\!-\epsilon m_0^2/128\kappa$ are the bending rigidity and saddle-splay coefficients of the square mean-curvature and Gaussian curvature, respectively \cite{Helf,BB}. The notation here is similar to that adopted by Blokhuis and Bedeaux \cite{BB}, although we have added a subscript to try to avoid confusion with the inverse bulk correlation length. Note there is no term proportional to the mean-curvature as required by the Ising symmetry.

A similar calculation reveals the general structure of the wall-$\beta$ interfacial free-energy. Consider the interface between a wall described by the height function $\psi({\bf{x}})$ and the bulk $\beta$ phase corresponding to spontaneous magnetization $m_0$. Recall that the magnetization at the surface $m_1$ is positive so that this interface does not exhibit any wetting behaviour. The DP result, Eq.\ (\ref{DPFwbeta}), for the excess free-energy involves only the area and local mean curvature of the wall. No higher order curvature corrections are present. Beyond DP approximation, we may reasonably expect this to change with the cubic and quartic interactions giving rise to additional curvature contributions. The perturbation theory is very similar to that described for the free interface and, to first-order, we find
\begin{equation}
F_{w\beta}[\psi]=F_{w\beta}^{(0)}[\psi]
+\epsilon\,\frac{\,\kappa m_0^2}{2}\Bigg\{t^3\fig{020}+\frac{\,t^4}{4}\fig{022}\!\!\Bigg\}+\;\cdots
\end{equation}
where this time the wavy line denotes the shape of the bounding wall. The diagrams are easily evaluated as an expansion in the inverse principal radii of curvature at the wall, and we find
\begin{eqnarray}
F_{w\beta}[\psi]=\int\!\!d{\bf{s}}_{\psi}\;
\Bigg\{\Sigma_{w\beta}+C_{w\beta}\left(\frac{1}{R_1^{\psi}}+\frac{1}{R_2^{\psi}}\right)\nonumber\\
+\frac{\kappa_{w\beta}}{2}\left(\frac{1}{R_1^{\psi}}+\frac{1}{R_2^{\psi}}\right)^2
+\frac{\bar\kappa_{w\beta}}{R_1^{\psi}R_1^{\psi}}+\;\cdots\Bigg\}
\label{Fwbetaexpan}
\end{eqnarray}
where the ellipses denote higher-order terms in the curvature. The new surface tension $\Sigma_{w\beta}$ and bending rigidity coefficient $C_{w\beta}$ contain very small corrections of order ${\mathcal O}(\epsilon\,t^3)$ to the DP results quoted earlier. The new rigidities $\kappa_{w\beta}\,\sim\,\bar\kappa_{w\beta}$ are ${\mathcal O}(\epsilon\,t^3)$ and are considerably smaller in magnitude than $C_{w\beta}$.

\section{First-order perturbation theory for $W[\ell,\psi]$}
\label{s5}

\subsection{General equations}

To begin, we restate the perturbation theory for the bulk potential in a slightly more general way. The calculation of the non-local binding potential only requires us to specify the form of the bulk potential in the wetting layer region where $m>0$. We write
\begin{equation}
\Delta \phi(m)= \frac{\,\kappa^2 \delta m^2}{2} +\sum_{n=3}^{\infty} \,\epsilon_n\,\kappa^2\, m_0^{2-n}\,\delta m^{n}
\label{PotPert2}
\end{equation}
where $\delta m=m-m_0$ and the $\epsilon_n$ are all dimensionless parameters. Thus, the usual "$m^4$" theory corresponds to $\epsilon_3=\frac{1}{2},\epsilon_4=\frac{1}{8}$ and $\epsilon_n=0$ for $n>4$. To first-order in perturbation theory, all the contributions are additive and we seek to write the non-local binding potential functional
\begin{equation}
W[\ell,\psi] = W^{(0)}[\ell,\psi]+ \sum_{n=3}^{\infty} \epsilon_n\, W^{(1)}_n [\ell,\psi]
+\;\cdots
\end{equation}
where, in an obvious notation, the $W_n^{(1)}$ are the perturbations corresponding to the term $\delta m^n$ in the bulk potential. To determine these, it is convienient to order the expansion of $\delta m^{(0)}_\Xi$ in the number of tubes that span the interfaces
\begin{eqnarray}
\delta m_{\Xi}^{(0)}=\left(\delta m_1\fig{014}-m_0\fig{011}\right)
\nonumber\\
-\left(\delta m_1\fig{015}-m_0\fig{012}\right)
\\ \nonumber
+\left(\delta m_1\fig{016}-m_0\fig{013}\right)-\;\cdots\hspace*{-1.cm}
\end{eqnarray}
 From (\ref{Pert1}), it follows that the first-order perturbations are given by
\begin{equation}
W_n^{(1)}[\ell,\psi]=\kappa^2\, m_0^{2-n}\int_{V_\beta}\!\! d{\bf{r}}\; \left(\delta m_{\Xi}^{(0)}\right)^n
- A_n^{(1)}[\ell] - B_n^{(1)}[\psi]
\label{Wn1}
\end{equation}
where $V_\beta$ denotes the volume of the wetting layer between the wall and interface. The functionals $A[\ell]$ and $B[\psi]$ do not describe interactions between the interface and wall and are introduced so that $W$ vanishes for infinite separation. For example,
\begin{equation}
A_4^{(1)}[\ell]=(-1)^4\kappa\, m_0^2\fig{023}
\end{equation}
\begin{equation}
B_4^{(1)}[\psi]=\kappa\, m_0^2\, t^4\fig{022}
\end{equation}
where, in each case, the wavy line denotes a configuration of the surface that corresponds to the argument of the functional. All that remains now is the evaluation of the integrals and the classification and simplification of the ensuing wetting diagrams.

\subsection{Wetting diagrams for Cubic and Quartic interactions}
\label{s42}

Substituting the magnetization profile into the first-order perturbation expression (\ref{Wn1}) for $n=3$ and $n=4$ leads to the following expressions for the first-order cubic and quartic corrections to the DP functional:

\begin{eqnarray}
\frac{W_{3}^{(1)}}{\kappa m_0^2}=\hspace*{6.cm}
\\ \nonumber
3t\left(\fig{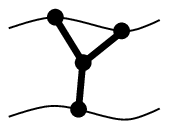}-\fig{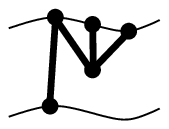}\right)-3t^2\left(\fig{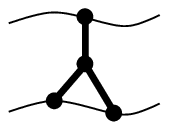}-\fig{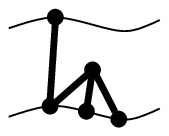}\right)
\\ \nonumber
+3\left(\fig{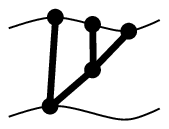}-\fig{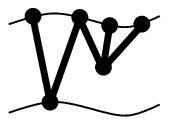}\right)-3t^3\left(\fig{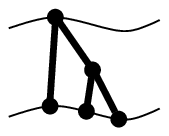}-\fig{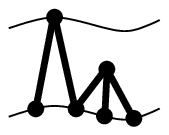}\right)
\\ \nonumber
3t\left(\fig{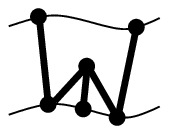}-2\fig{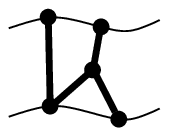}\right)-3t^2\left(\fig{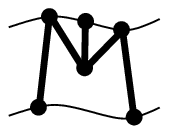}-2\fig{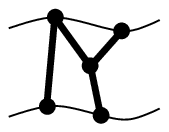}\right)
\end{eqnarray}
and
\begin{eqnarray}
\frac{W_{4}^{(1)}}{\kappa m_0^2}=\hspace*{6.cm}
\\ \nonumber
-4t\left(\fig{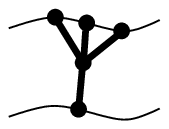}-\fig{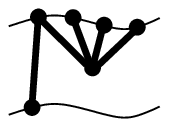}\right)-4t^3\left(\fig{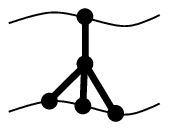}-\fig{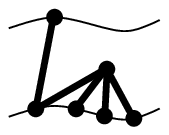}\right)
\\ \nonumber
-4\left(\fig{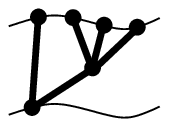}-\fig{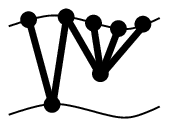}\right)-4t^4\left(\fig{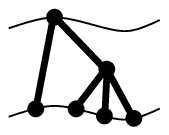}-\fig{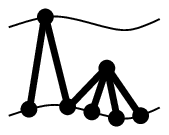}\right)
\\ \nonumber
6t^2\left(\fig{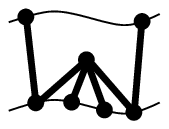}-2\fig{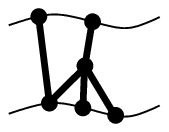}\right)+6t^2\left(\fig{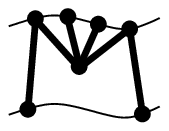}-2\fig{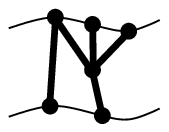}\right)
\\ \nonumber
+6t^2\fig{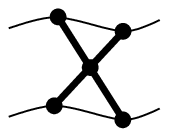}\hspace*{5.5cm}
\end{eqnarray}

Higher-order diagrams exist but involve at least three tubes that span the surfaces and would generate terms of order ${\mathcal O}(e^{-3\kappa\ell})$ in the standard binding potential function. Each of the new wetting diagrams has one black dot lying between the surfaces and represents an integral over the volume $V_\beta$ of the wetting layer. The associated infinitesmal measure is $\kappa d{\bf{r}}$. Thus, the first wetting diagram in the expansion of $W_{3}^{(1)}$ is
\begin{eqnarray}
\fig{029}=\hspace*{4cm}
\\ \nonumber
\kappa\int\!\! d{\bf{s}}_{\psi}\!\int_{V_\beta}\!\! d{\bf{r}}\;K({\bf{r}}_{\psi},{\bf{r}})\,
\left\{\int\!\!d{\bf{s}}_{\ell}\;K({\bf{r}},{\bf{r}}_{\ell})\right\}^2
\end{eqnarray}
where we have labelled the points in an obvious notation. It is natural to interpret this as a {\it{splitting}} of a tube-like fluctuation connecting the surfaces. The second diagram in the same cubic interaction does not involve a splitting but instead adds a "corona" corresponding to short tube-like fluctuations away from the interface:
\begin{eqnarray}
\fig{030}=\hspace*{5.cm}
\\ \nonumber
\kappa\!\int\!\!\!\!\int\!\!\!\!\int_{V_\beta}\!\!d{\bf{s}}_{\psi}d{\bf{s}}'_{\ell}d{\bf{r}}\;
K({\bf{r}}_{\psi},{\bf{r}}'_{\ell})\,K({\bf{r}}'_{\ell},{\bf{r}})\,
\left\{\int\!\!d{\bf{s}}_{\ell}\;K({\bf{r}},{\bf{r}}_{\ell})\right\}^2\!\!.
\end{eqnarray}
Similar intepretations apply to all the wetting diagrams. One contribution which is of particular novelty is the $\mathcal{X}$ diagram
\begin{eqnarray}
\fig{006}=\hspace*{6.cm}
\\ \nonumber
\kappa\!\int\!\!\!\!\int\!\!\!\!\int\!\!\!\!\int\!\!\!\!\int_{V_\beta}\!\!d{\bf{s}}_{\psi}d{\bf{s}}'_{\psi}
d{\bf{s}}_{\ell}d{\bf{s}}'_{\ell}d{\bf{r}}\;
K({\bf{r}}_{\psi},{\bf{r}})\,
K({\bf{r}}'_{\psi},{\bf{r}})\,
K({\bf{r}},{\bf{r}}_{\ell})\,
K({\bf{r}},{\bf{r}}'_{\ell})
\label{Xfirst}
\end{eqnarray}
and arises from the quartic interaction. This has an appealling physical intepretation as a local pinching of two tubes that span the surfaces. As we shall see, this is a rather interesting diagram even though ultimately it does not influence the leading-order physics.

\subsection{Wetting diagram relations}
\label{s43}

The cubic and quartic interactions appear to give rise to a plethora of new wetting diagrams. However, the physics represented by these perturbations is rather simple and can be elegantly expressed in a more concise fashion. The essential ingredients in this simplification are various relations between the wetting diagrams which express their reducibility. We will illustrate this with some examples.

Consider the first wetting diagram appearing in $W_3^{(1)}$. To begin, suppose that the wetting layer has planar area $A_w$ and is of constant thickness $\ell$. The integrals are easily evaluated yielding
\begin{equation}
\fig{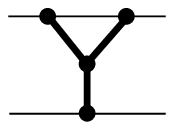}=A_w (e^{-\kappa\ell}-e^{-2\kappa\ell})
\end{equation}
This can be expressed diagrammatically
\begin{equation}
\fig{053}=\fig{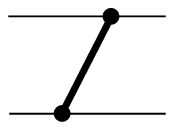}-\fig{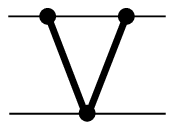}
\end{equation}
showing the perturbative diagram is reducible to the DP contributions $\Omega_1^1$ and $\Omega_1^2$. The net effect of this diagram is, therefore, to simply shift the coefficients 
\begin{eqnarray}
a^{(0)}_1\;\rightarrow\;a_1=a^{(0)}_1+3\epsilon_3\,t\,\kappa\,m_0^2\\
b^{(0)}_1\;\rightarrow\;b_1=b^{(0)}_1-3\epsilon_3\,t\,\kappa\,m_0^2
\end{eqnarray}
appearing in the DP expression for $W$. Moreover, a nice feature of the perturbation theory is that there is no need to keep precise book-keeping concerning such shifts. This can be done exactly at the end of the calculation once the general diagrammatic structure has been elucidated.

The above expression is not quite the whole relation for the wetting diagram since interfacial and substrate curvature are not allowed for. More generally, one finds (after a few integrations)
\begin{equation}
\fig{029}=\fig{001}+\frac{1}{2}\fig{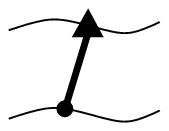}-\fig{004}+\;\cdots
\label{wettingdiagram1}
\end{equation}
where we have introduced a new type of diagram containing a black triangle. The triangle will always lie on a surface and is intepreted as an integral over the surface with local measure $d{\bf{s}}$ multiplied by the sum of the local
principal curvatures, measured in units of $\kappa$ (to preserve the units of the diagrams). Thus,
\begin{equation}
\fig{002}=\;\frac{1}{\kappa}\int\!\!\!\!\int\!\! d{\bf{s}}_{\psi}d{\bf{s}}_{\ell}\;
K({\bf{r}}_{\psi},{\bf{r}}_{\ell})\left(\frac{1}{R_1^{\ell}}+\frac{1}{R_2^{\ell}}\right)
\label{75}
\end{equation}
and similarly if a triangle is placed on the wall. The ellipses in the wetting diagram relation ({\ref{wettingdiagram1}) denote higher order curvature terms which are negligible.

Similarly, for the second wetting diagram in $W_3^{(1)}$, one can write the relation
\begin{eqnarray}
\fig{030}=\;\frac{1}{3}\,\fig{001}+\frac{1}{18}\fig{002}+\;\cdots
\end{eqnarray}
where here the ellipses also include terms involving four tubes that span the surfaces as well as higher-order curvatures. The same process is also valid for diagrams with two tubes spanning the surfaces. For example

\begin{eqnarray}
\fig{034}=\;\frac{1}{3}\,\fig{004}+\;\cdots
\end{eqnarray}
Again the effect of these diagrams is to shift the coefficient of the $\Omega_1^2$ diagram and add negligible curvature terms. In the first-order perturbation theory all bar one diagram can be recast as a sum of the DP diagrams $\Omega_1^1$, $\Omega_1^2$ and $\Omega_2^1$ together with curvature corrections. The only contribution for which there is no such relation is the $\mathcal{X}$ diagram describing the two-tube pinching process (\ref{Xfirst}) which is not reducible. However, relations involving it do emerge at second-order in perturbation theory.

In summary, three effects emerge at first-order in perturbation theory: 1) Rescaling of the coefficients $a_1$, $b_1$, etc.\ 2) appearance of curvature corrections and 3) introduction of non-zig-zag diagrams describing tube interactions.

\section{Second-order perturbation theory for $W$}
\label{s6}

At second-order, by far the most important contribution arises from the cubic interaction in $\Delta\Phi^{(1)}$. Contributions of order $\epsilon_4^2$, as well as mixing terms $\epsilon_3\epsilon_4$, are small and do not introduce any new physics. For ease of presentation, we suppose that the potential perturbation has only one power, $\Delta\Phi^{(1)}=\kappa^2\, m_0^{2-n}\,\delta m^n$, and determine the second-order term in
\begin{equation}
W[\ell,\psi]=W^{(0)}[\ell,\psi]+\epsilon_n\, W^{(1)}[\ell,\psi]+\epsilon^{2}_n\, W^{(2)}[\ell,\psi]
\end{equation}
Setting $n=3$ at the end of the calculation reveals the dominance of the cubic interaction at this order. The second-order perturbation is
\begin{equation}
W^{(2)}[\ell,\psi]=\frac{\,n\,\kappa^2 m_0^{2-n}}{2}\int\!\! d{\bf{r}} \;\delta m^{(1)}_{\Xi}
(\delta m^{(0)}_{\Xi})^{n-1}\,-A_n^{(2)}[\ell]-B_n^{(2)}[\psi]
\label{Wpert2}
\end{equation}
where, as in the first-order perturbation theory, functionals $A_n^{(2)}[\ell]$ and $B_n^{(2)}[\psi]$ are introduced so that, by construction, $W^{(2)}$ vanishes when the interface is delocalised from the wall. They need not be specified explicitly, as they are automatically generated by the integral in (\ref{Wpert2}).

The second-order term in the potential $W$ depends on the first-order correction to the profile $\delta m^{(1)}_{\Xi}$ which satisfies
\begin{equation}
 \nabla^{2}\delta m^{(1)}_{\Xi}=\kappa^2\,\delta m^{(1)}_{\Xi}+n\,\kappa^2\,m_0^{2-n}\,(\delta m^{(0)}_{\Xi})^{n-1}
\end{equation}
Substitution of the DP profile leads to the PDE
\begin{eqnarray}
\nabla^{2}\delta m^{(1)}_{\Xi}=\kappa^2\,\delta m^{(1)}_{\Xi}\hspace*{5.cm}
\nonumber\\[.25cm]
-n(-1)^n\kappa^2 m_0\,\Bigg\{(n-1)\,t\,\Bigg(\fig{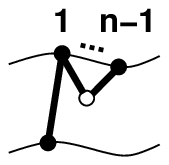}-\fig{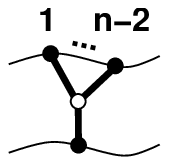}\Bigg)
\nonumber\\
+\fig{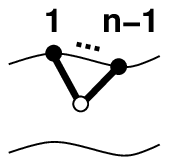}+(n-1)\Bigg(\fig{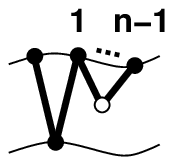}-\fig{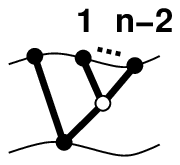}\Bigg)\Bigg\}\hspace*{.5cm}
\end{eqnarray}
where we have curtailed the expansion at two tubes spanning the surfaces, and neglected terms of ${\mathcal O}(t^2)$. The inhomogeneous PDE can be solved in a standard manner using the same Green's function $K({\bf{r}}_1,{\bf{r}}_2)$. Thus, the solution can also be written diagrammatically and, after some algebra, we find
\begin{eqnarray}
\delta m^{(1)}_{\Xi}=\frac{(-1)^n n}{2}\Bigg\{
(n\!-\!1)\,\delta m_1\Big[\left(\fig{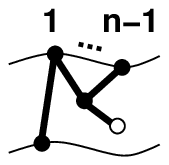}\!-\!\fig{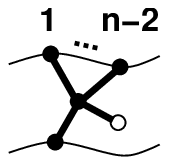}\right)\hspace*{1.5cm}
\nonumber\\[.2cm]
-\left(\fig{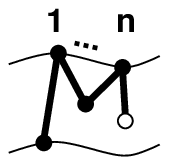}\!-\!\fig{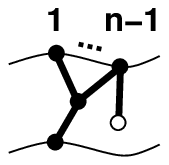}\right)\Big]
+m_0\Big[\left(\fig{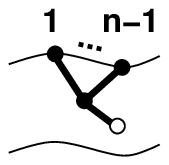}\!-\!\fig{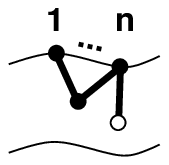}\right)\hspace*{1.5cm}
\nonumber\\[.2cm]
+\left(\fig{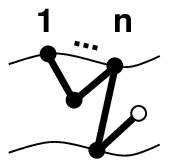}\!-\!\fig{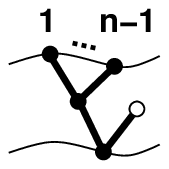}\right)-\left(\fig{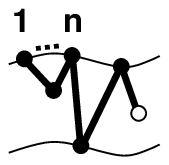}\!-\!\fig{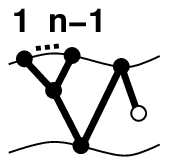}\right)\Big]\hspace*{1.5cm}
\nonumber\\[.2cm]
+\;(n\!-\!1)\left(\fig{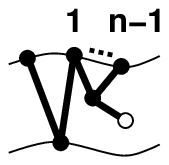}\!-\!\fig{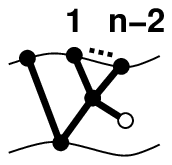}\right)\hspace*{3.25cm}
\nonumber\\[.2cm]
-\;(n\!-\!1)\left(\fig{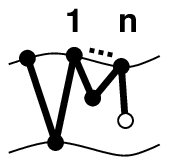}\!-\!\fig{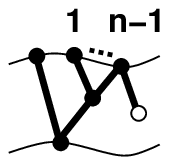}\right)
\Bigg\}\hspace*{3.cm}
\nonumber\\[-1.cm]
\end{eqnarray}

Specialising in the dominant cubic interaction $(n=3)$, we find for the second-order perturbation in $W$:
\begin{equation}
\frac{\,W_3^{(2)}[\ell,\psi]}{\kappa\, m_0^2}=\,-\frac{\,9}{4}\left\{\,4\,t\,D_1^1+ D_1^2\right\}+{\mathcal O}(t^2)
\label{D}
\end{equation}
where the $D_1^1$ and $D_1^2$ denote the following sum of diagrams:
\begin{eqnarray}
D_1^1=
\fig{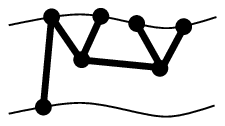}\;\;-\!\fig{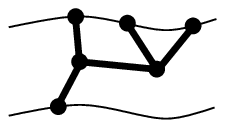}
\nonumber\\
-\fig{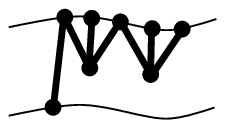}\;\;+\!\fig{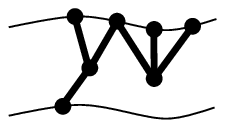}
\end{eqnarray}
and
\begin{eqnarray}
D_1^2=2\fig{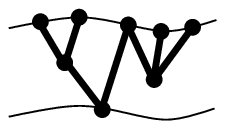}
\\
-\fig{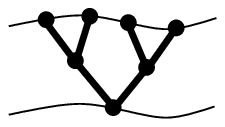}-\fig{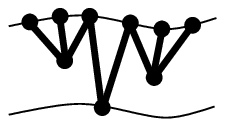}
\nonumber\\
+4\fig{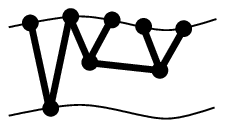}-4\fig{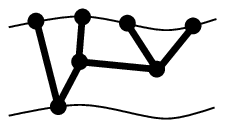}
\nonumber\\
+4\fig{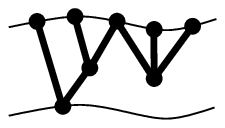}-4\fig{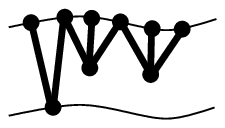}
\end{eqnarray}

These diagrams determine the rescaling of the coefficients $a_1$ and $b_1$, and also generate curvature corrections due to the interface.  Again, the key to understanding their net effect is through wetting diagram relations. For example, the following quartic diagram can be expressed
\begin{equation}
\fig{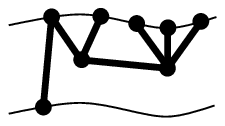}\;\;\;=\;\;\;\frac{1}{3}\fig{001}+\frac{2}{9}\fig{002}+\;\cdots
\end{equation}
where the ellipses include higher-order interfacial curvature terms and four-tube diagrams. In this way, each of the contributions in (\ref{D}) can be written as a sum of the diagrams
\begin{eqnarray}
\fig{001},\qquad\fig{002},\qquad\fig{004}
\end{eqnarray}
similar to the first-order perturbation theory. If one extends the calculation to allow terms of order $t^2$, $t^3$, etc.\ , one also encounters wetting diagrams where corona-like tubes eminate from the substrate. These are, in fact, the same as the diagrams in $D_1^1$ and $D_1^2$ but with the interfacial and substrate surfaces switched. Thus, for example,
\begin{eqnarray}
\fig{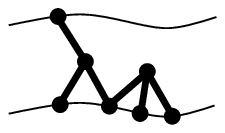}
\end{eqnarray}
has a coefficient proportional to $t^3$ and will add higher-order powers of $t$ in the expansion of $a_1$, and also generate curvature corrections due to the substrate which can be recast in terms of the diagram
\begin{eqnarray}
\fig{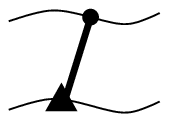}=\frac{1}{\kappa}\int\!\!\!\!\int\!\!d{\bf s}_\psi d{\bf s}_\ell\;
K(r_\psi,r_\ell)\,\left(\frac{1}{R_1^\psi}+\frac{1}{R_2^\psi}\right).
\label{90}
\end{eqnarray}
Again, the general structure obtained from the first-order perturbation theory is unchanged.

Working to ${\mathcal O}(t^2)$, one also generates wetting diagrams which are closely related to the two-tube pinching process which arose in the first-order perturbation from the
quartic interaction. For example,
\begin{eqnarray}
\fig{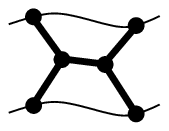},
\label{XXY}
\end{eqnarray}
whose coefficient is proportional to $t^2$. The two central black dots in the wetting layer are connected by a tube-like fluctuation which does not attach to either the wall or interface. The connecting tube is necessarily of short length because the corresponding integral is heavily damped by the Kernel $K$. This is neatly expressed diagrammatically
\begin{eqnarray}
\fig{099}=\;2\fig{006}-\fig{004}-\fig{005}+\;\cdots
\end{eqnarray}
leading to the rescaling of the coefficients of $\Omega_1^2$, $\Omega_2^1$ and ${\mathcal X}$. Curvature corrections, represented by the ellipsis, are of negligible importance for two-tube diagrams.

In summary, second-order perturbation theory leads to the same three effects highlighted in the first-order calculation: the rescaling of coefficients, and the appearance of curvature and tube-interaction diagrams.

\section{The general structure of the non-local binding potential.}
\label{s7}

The general structure of the non-local binding potential functional for short-ranged wetting is now apparent. Up to "two tubes", the functional has an asymptotic large distance decay described by the diagrams
\begin{eqnarray}
W=a_1\fig{001}+c_1\fig{002}+c_2\fig{003}
\nonumber\\
+b_1\fig{004}+b_2\fig{005}+d_1\fig{006}+\;\cdots\qquad
\label{asym}
\end{eqnarray}
which should be compared with the DP result (\ref{27}). Thus, going beyond DP generates curvature terms (shown for one-tube diagrams only) and also a tube-interaction diagram. The corresponding algebraic expressions are given by equations (\ref{24}),(\ref{75}),(\ref{90}),(\ref{25}),(\ref{26}) and (\ref{Xfirst}). The coefficients are geometry independent and all have power series expansions in the scaling field $t$. The leading-order behaviours are
\begin{eqnarray}
\frac{a_1}{\kappa m_0^2}=\alpha t,\qquad\frac{b_1}{\kappa m_0^2}=\beta,
\qquad\frac{b_2}{\kappa m_0^2}=\beta t^2,\nonumber\\
\frac{c_1}{\kappa m_0^2}=\gamma t,\qquad\frac{c_2}{\kappa m_0^2}=\gamma
t^2,\qquad\frac{d_1}{\kappa m_0^2}=\chi t^2
\label{abcd}
\end{eqnarray}
and are specified by just four dimensionless constants reflecting the surface exchange symmetry of $W$. The coefficients $b_2$, $c_2$ and $d_1$ all vanish as $t^2$ implying that the associated diagrams are of negligible importance at critical wetting. The second diagram, describing the curvature correction due to the $\alpha |\beta$ interface, is necessarily much smaller than $\Omega_1^1$ and is therefore also negligible given that $c_1$ also vanishes at the critical wetting phase boundary. Thus, the diagrammatic expression for $W$ is the same as calculated using the DP approximation in \cite{NoLo} but with different numerical coefficients. This is the main finding of our study.

The exact values of the above coefficients can be calculated for the "$m^4$" LGW potential (\ref{m4}), by matching with mean-field results for specific interfacial and wall configurations. Consider for example the simplest situation of a flat wall, $\psi=0$ and a flat interface $\ell({\bf{x}})=\ell$. The corresponding planar binding potential function is defined as
\begin{equation}
W_{\pi}(\ell)=\frac{W[\ell,0]}{A_w}\Bigg|_{\ell({\bf{x}})=\ell}
\end{equation}
and can be identified with the diagrams
\begin{eqnarray}
A_w\,W_{\pi}(\ell)=a_1\fig{054}+b_1\fig{055}
\nonumber\\
+b_2\fig{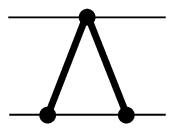}+d_1\fig{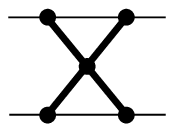}+\;\cdots
\end{eqnarray}
The first three diagrams are of DP-type and were discussed in \cite{NoLo}. The new diagram can also be evaluated exactly
\begin{equation}
\fig{105}=\,A_w \, \kappa \ell\, e^{-2\kappa \ell}
\end{equation}
implying that there are non-purely exponential terms in the binding potential. Thus, the binding potential function necessarily has the general expansion
\begin{equation}
W_{\pi}(\ell)=a_1e^{-\kappa \ell}+(b_1+b_2+d_1 \kappa\ell) e^{-2\kappa \ell}+\;\cdots
\label{Wpi}
\end{equation}
with coefficients specified in (\ref{abcd}). This is identical to the findings of Fisher and Jin who calculated $W_{\pi}(\ell)$ directly \cite{FJ2}. One advantage of the Green's function approach is that the diagram leading to the non-purely exponential term is isolated and can be evaluated for other geometries. For example, for spherical interfacial and wall shapes
\begin{equation}
\fig{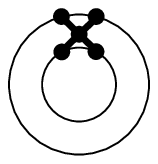}=\sqrt{A_w\,A_{\alpha\beta}}\;\kappa\,\ell\;e^{\,-2\kappa\ell}
\end{equation}
where, as in \cite{NoLo}, $A_w=4\pi\,R^2$ and $A_{\alpha\beta}=4\pi\,(R+\ell)^2$ are the areas of the wall and interfacial configurations, respectively.

We can now determine the coefficients $a_1$, $b_1$, $\dots$ by comparing (\ref{Wpi}) with the known asymptotic decay of $W_{\pi}$ for arbitrary potentials $\Delta\Phi(m)$. This can be calculated independently without recourse to perturbation theory. For planar interfacial and wall configurations, the constrained profile $m_{\Xi}\equiv m_{\pi}(z;\ell)$ satisfies the "energy-conservation" condition
\begin{equation}
\frac{1}{2}\Bigg(\frac{\partial m_{\pi}}{\partial z} \Bigg)^2=
\Delta \phi (m_{\pi})- W_{\pi}'({\ell})
\label{quad}
\end{equation}
This can be integrated, and the large distance expansion exactly determined. For the "$m^4$" potential (\ref{m4}), we find
\begin{eqnarray}
\frac{a_1}{\kappa m_0^2}=4t,\qquad\quad\frac{b_1}{\kappa m_0^2}=4\hspace*{.4cm}
\nonumber\\[.2cm]
\frac{b_2}{\kappa m_0^2}=4t^2,\qquad\quad\frac{d_1}{\kappa
m_0^2}= 6t^2
\end{eqnarray}
The curvature coefficient $\gamma=-8$ can be determined in a similar fashion by considering spherical wall and interfacial configurations.

One can go further in this analysis and determine the coefficients for the perturbative potential (\ref{phi0+phi1}) to all orders in $\epsilon$. We only quote the results for $a_1$ and $b_1$
\begin{equation}
\frac{a_1}{\kappa m_0^2}=\frac{8t}{2-\epsilon+\sqrt{4-3\epsilon}},\qquad
\frac{b_1}{\kappa m_0^2}=\frac{16}{(2-\epsilon+\sqrt{4-3\epsilon})^2}
\end{equation}
which smoothly interpolate between the DP and "$m^4$" theory results.

\section{Non-local model for wetting at planar substrates}
\label{s8}

In this section, we show how all the wetting diagrams appearing in the asymptotic expansion (\ref{asym}) simplify when the substrate is planar ($\psi=0$). Clearly, there is no contribution from substrate curvature and we write the interfacial model
\begin{equation}
H[\ell]=H_{\alpha\beta}[\ell]+ W[\ell]
\end{equation}
with planar binding potential functional ($W[\ell]\equiv W[\ell,0]$)
\begin{eqnarray}
W[\ell]=a_1\fig{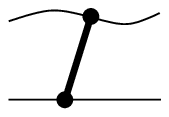}+c_1\fig{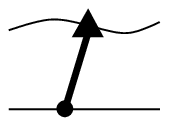}\hspace*{3.cm}
\nonumber\\
+\;b_1\fig{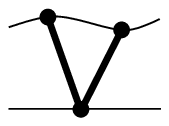}+b_2\fig{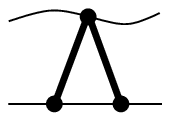}+d_1\fig{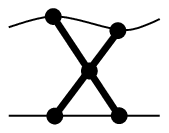}+\;\cdots
\end{eqnarray}
containing two new diagrams compared to the corresponding DP expression. Three of these diagrams can be evaluated by simply holding the dot (or triangle) on the upper interface fixed and integrating over the wall:
\begin{equation}
\fig{110}=\int d{\bf{x}}\,\sqrt{1+(\nabla\ell)^2}\;e^{-\kappa \ell},
\label{omega11flat}
\end{equation}
\begin{equation}
\fig{111}=\int d{\bf{x}}\;\sqrt{1+ (\nabla l)^2}\,\left(\frac{1}{R_1^{\ell}}+\frac{1}{R_2^{\ell}}\right)
e^{-\kappa\ell}
\label{111}
\end{equation}
and
\begin{equation}
\fig{112}=\int d{\bf{x}} \sqrt{1+(\nabla\ell)^2}e^{-2\kappa \ell}
\label{omega21flat}
\end{equation}
which are all local contributions to the effective Hamiltonian $H[\ell]$. In particular, if $\nabla \ell$ is small, one can see how each contribute towards a local binding potential function and/or effective position-dependent stiffness.
Note that, if the coefficient $c_2$ of the curvature diagram (\ref{111}) is zero, as it is at DP level, the $\Omega_1^1$ diagram (\ref{omega11flat}) determines the leading order exponential decays of $W_\pi(\ell)$ and $\Sigma(\ell)$. Beyond DP, however, the curvature diagram also contributes to $\Sigma(\ell)$.

In contrast, the two remaining diagrams, $\Omega_1^2$ and ${\mathcal{X}}$, are strongly non-local. As remarked in \cite{PRL} application of the convolution theorem reduces the triple integral (\ref{25}) to a double integral
\begin{equation}
\fig{007}=\int\!\!\!\!\int\!\! d{\bf{s}}_1 d{\bf{s}}_2\;\,
e^{-\kappa\ell({\bf{x}}_1)}
S(x_{12};\bar{\ell}_{12})\;
e^{-\kappa\ell({\bf{x}}_2)}
\label{twobody1}
\end{equation}
where $\bar{\ell}_{12}=(\ell({\bf{x}}_1)+\ell({\bf{x}}_2))/2$ is the mean interfacial height of the two points at the interface. Here $S$ is a two-body interfacial interaction which decays as a two-dimensional Gaussian
\begin{equation}
S(x_{12};\ell)\approx\frac{\kappa}{4\pi\ell} \exp\left(-\frac{\kappa x_{12}^2}{4 \ell}\right)
\label{twobody2}
\end{equation}
and which controls the repulsion of the interface from the wall. By construction, the integrated strength of $S$ is unity. There are two features about this effective many-body interaction which are worth commenting on. Firstly, its range increases as the square-root of the film thickness and, therefore, becomes longer ranged as the interface unbinds. It is this that necessitates a non-local treatment of short-ranged critical wetting, and is responsible for the breakdown of local theories. Also, the same Gaussian interaction (\ref{twobody2}) follows from a simple saddle-point evaluation of the integral (\ref{25}) over the wall. This means that the interaction between two fixed points on the interface arises due to a connecting tube that reflects off the wall and is of minimal length. This physical interpretation will be useful in discussions of wetting at non-planar walls, where an exact convolution evaluation of $\Omega_{1}^2$ is not available.

Similar arguments apply to the ${\mathcal{X}}$ diagram, describing the two-tube interaction which can be written
\begin{equation}
\fig{008}=\int\!\!\!\!\int\!\! d{\bf{s}}_1 d{\bf{s}}_2\;\,
e^{-\kappa\ell({\bf{x}}_1)}
X(x_{12};\bar{\ell}_{12})\;
e^{-\kappa\ell({\bf{x}}_2)}
\label{twobodyX}
\end{equation}
The two-body interaction describing this interaction also depends on the mean-interfacial height only, and is given by
\begin{equation}
X(x;\ell)=\frac{\,\kappa^2}{4\pi}\,\Gamma\Big(0\,,\frac{\kappa x^2}{4\ell}\Big)
\end{equation}
where $\Gamma(0,z)$ is the incomplete gamma function. At large distances, this decays similar to the two dimensional Gaussian (\ref{twobody2}).

Finally, we mention that in the strict small gradient limit the non-local Hamiltonian reduces to
\begin{equation}
H[\ell]=\int\!\!d{\bf{x}}\;\,\Bigg\{\frac{\Sigma(\ell)}{2}(\nabla\ell)^2+W_{\pi}(\ell)\Bigg\}+\Sigma_{\alpha\beta}\,A_w
\end{equation}
where the position dependent contributions to the binding potential and stiffness coefficient have the general decays
\begin{equation}
W(\ell)=\,w_{10}\,e^{-\kappa\ell}+(w_{21}\kappa\ell+w_{20})\,e^{-2\kappa\ell}+\;\cdots
\end{equation}
and
\begin{equation}
\Delta\Sigma(\ell)=\,s_{10}\,e^{-\kappa\ell}+
(s_{22}\,\kappa^2\ell^2+s_{21}\,\kappa\ell+s_{20})\,e^{-2\kappa\ell}+\;\cdots
\end{equation}
respectively. All seven coefficients exhibit power-law dependences on the scaling field $t$ and are determined by the five coefficients $a_1$, $b_1$, $c_1$, $b_2$ and $d_1$. We find $w_{10}\sim s_{10}\sim t$, $w_{21}\sim s_{22}\sim t^2$ and all other coefficients finite at $t=0$. These are in precise agreement with the local theory of Jin and Fisher \cite{FJ1}.

\section{Discussion}
\label{s9}

In this paper, we have extended our earlier derivation of a non-local interfacial Hamiltonian for short-ranged wetting beyond the DP approximation. We have shown that the diagrammatic method introduced in our first paper combines rather nicely with perturbation theory and allows us to derive the general structure of the binding potential functional $W$. While this contains some new diagrams describing curvature corrections and tube-interations, these are of negligible importance for wetting. The dominant diagrammatic structure of $W$ is the same as that derived for the DP model, albeit with slightly shifted coefficients. The values of these coefficients have been determined exactly. We also showed how all the diagrams in the asymptotic expansion of $W$ (up to two tubes) simplify for wetting at planar substrates. While some local contributions are accurately described by a binding potential function and position dependent stiffness, the non-local contributions generate weak long-ranged two-body interfacial interactions, which play a crucial role at critical wetting.

To finish our article, we make a number of remarks concerning the interpretation, limitations, and further extension of the present approach:

1.- {\sl New features}. Going beyond DP approximation alters the binding potential functional in three different ways: rescaling of the coefficients $a_1$, $b_1\,\dots$, and the introduction of curvature and tube interaction diagrams. These effects could have been anticipated on very general grounds. Even for a free Hamiltonian, going beyond DP alters the surface tension and introduces Helfrich-like rigidity terms, consistent with morphological expectations. The first two effects merely mirror those in $W$. Also, there must be a mechanism which generates non purely exponential terms in the binding potential function $W_\pi(\ell)$ beyond DP. This is fulfilled by the two-tube interaction diagram $\mathcal X$.

2.- {\sl Generalisation of the LGW model}. One could consider a slightly generalised LGW model in which the coefficient of $(\nabla m)^2$ is a function of the order-parameter \cite{RW}. By expanding this function about $m_0$, it is straightforward to show that the corresponding first-order perturbation corrections to $W$ are equivalent to cubic, quartic, etc.\ corrections to the DP potential, which we have considered explicitly. The same is also true of square-Laplacian contributions to $H[m]$ which are equivalent to a shift in the value of $\kappa$ plus the aformentioned cubic, quartic and higher-order corrections. In both cases, the diagrammatic structure of $W$ remains unaltered to first-order. Of course, this robustness is to be expected. As remarked at the end of our first paper, the form of $W$ is necessitated by exact sum-rule requirements.

3.- {\sl Coupling to a surface field and enhancement}. While we have restricted ourselves to fixed boundary conditions at the wall, it is straightforward to extend the Green's function method to different kinds of surface boundary conditions. In particular, one may consider LGW models of the form \cite{NF}
\begin{equation}
H[m]=\int\!\!d{\bf r}\;\left\{\,\frac{1}{2}(\nabla m)^2+\Delta\phi(m)\,\right\}\;+\;
\int\!\!d{\bf s}_\psi\;\phi_1(m({\bf r}_\psi))
\end{equation}
where $\phi_1(m)=-\frac{g}{2}m^2-h_1\,m$ describes the coupling to a surface field $h_1$ and enhancement $g$. In this case, the divergence theorem generalises eq.\ (\ref{div}) to 
\begin{eqnarray}
H^{(0)}[m_{\Xi}^{(0)}]=-\frac{1}{2}\int_{\psi}d{\bf{s}}_{\psi}\,(m_{\Xi}^{(0)}-m_0)\,\phi_1'(m_{\Xi}^{(0)})
\\ \nonumber
-\frac{ m_0}{2}\int_{\ell^-}d{\bf{s}}_{\ell}\nabla
m_{\Xi}^{(0)}\cdot{\bf{n}}_{\ell}-
\frac{m_0}{2}\int_{\ell^+}d{\bf{s}}_{\ell}\nabla
m_{\Xi}^{(0)}\cdot{\bf{n}}_{\ell}
\end{eqnarray}
where we have used the appropriate boundary condition in the integral over the wall. Similarly, the diagrammatic expansion for the constrained magnetisation reads, in DP approximation,
\begin{eqnarray}
\delta m_{\Xi}^{(0)}=-m_0\left(\fig{011}-\fig{012}+\fig{013}-\;\cdots\right)
\nonumber\\
+\left(\fig{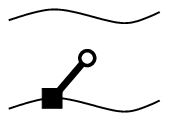}-\fig{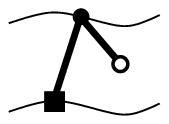}+\fig{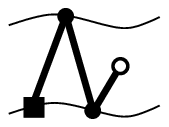}-\;\cdots\right)\qquad\quad
\end{eqnarray}
where the black squares on the wall denotes a convolution of the Green's function $K$ with an auxiliary function $\mu({\bf r}_\psi)$. This accounts for the variation of the surface magnetisation, and is introduced to satisfy the boundary condition at the wall. For example, for a planar wall, $\mu$ satisfies the integral equation
\begin{eqnarray}
\kappa\,\mu({\bf r}_\psi)+2\,\kappa\,m_0\,\fig{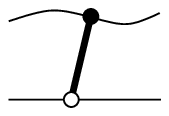}=\hspace*{3.cm}
\\ \nonumber
g\,m_0+h_1+g\int\!\!d{\bf r}'_\psi\;\,\mu({\bf r}'_\psi)\;K\big({\bf r}'_\psi,{\bf r}_\psi\big)
\end{eqnarray}
which can be solved via a Fourier transform. From here, the calculation proceeds as described in \cite{NoLo}, within DP approximation, and similar to that described herein using perturbation theory. The final result for the two dominant diagrams in $W$ remains unchanged from eq.\ (\ref{1}) but with coefficients
\begin{eqnarray}
a_1\,=\,\frac{\kappa}{\kappa-g}\;\frac{\;8\,(h_1 m_0+g\,m_0^2)\;}{2-\epsilon+\sqrt{4-3\epsilon\,}}\hspace*{.4cm}
\nonumber \\ \\ \nonumber
b_1\,=\,-\frac{\kappa+g}{\kappa-g}\;\frac{16\,\kappa\,m_0^2}{(2-\epsilon+\sqrt{4-3\epsilon})^2}
\end{eqnarray}
where we have expressed the results as appropriate for the potential (\ref{phi0+phi1}). As anticipated, the diagrammatic form is very similar to that described for fixed boundary conditions but with the advantage that one can now discuss first-order wetting ($g>-\kappa$) and tricritical wetting ($g=-\kappa$). The problem of tricritical wetting is particularly interesting because the coefficient $b_1$ of the second diagram, $\Omega_1^2$, in (\ref{1}) vanishes. The repulsion from the wall is then determined by a diagram which generates a term of order $e^{\,-3\,\kappa\ell}$ in the usual binding potential function. This corresponds to the next diagram in the series (\ref{asym}) and involves three tubes, as is discussed next.

4.- {\sl The dominant three-tube diagram}. There are several diagrams involving three tubes which contribute towards the coefficient of $e^{\,-3\,\kappa\ell}$ in $W_\pi(\ell)$. However, there is only one such diagram whose coefficient does not vanish at $t=0$ and is, therefore, necessary for the discussion of tricriticality. The diagram in question is
\begin{eqnarray}
\fig{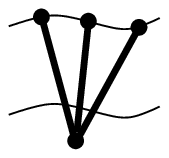}=\kappa\int_{V_w}\! d{\bf{r}}\; \left\{\int\!\!d{\bf{s}}_{\ell}\; K({\bf{r}}_{\ell},{\bf{r}})\,\right\}^3
\end{eqnarray}
and is generated by the cubic interaction in $\Delta\phi(m)$. Here, $V_w$  denotes the volume of the wall. This diagram is the next term in the asymptotic expansion (\ref{asym}) and is strongly non-local. However, analogous to our previous discussion of $\Omega_1^2$, the diagram simplifies and generates an effective many-body interfacial interaction between points at the interface. For example, for a planar wall, the integral reduces to
\begin{eqnarray}
\fig{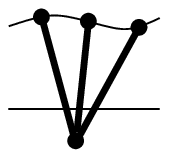}=\hspace*{6.cm}
\\
\nonumber
\int\!\!\!\!\int\!\!\!\!\int\!\!d{\bf s}_1\,d{\bf s}_2\,d{\bf s}_3\;
e^{-\kappa\ell({\bf x}_1)}\,e^{-\kappa\ell({\bf x}_2)}\,e^{-\kappa\ell({\bf x}_3)}\;T(x_{12},x_{23},x_{13})
\end{eqnarray}
where the three-body interaction
\begin{equation}
T(x_{12},x_{23},x_{13})\propto
\lambda_{123}\,\exp\!\left(-\lambda_{123}\,(x_{12}^2\ell_3+x_{23}^2\ell_1+x_{13}^2\ell_2)\right)
\end{equation}
Here,
\begin{eqnarray}
\lambda_{123}\,=\;\frac{\kappa}{2(\ell_1\ell_2+\ell_2\ell_3+\ell_1\ell_3)}
\end{eqnarray}
and we have abbreviated $\ell_1=\ell({\bf x}_1)$, etc.

The fluctuation theory of three-dimensional short-ranged tricritical wetting requires a renormalisation-group treatment of the flow of this three-body interaction similar to that described in \cite{PRL} for the two-body term $S$, eq.\ (\ref{twobody2}), pertinent to critical wetting. We note that the range of this three-body interaction also increases as the interface depins implying that non-local effects are important at this transition.

5.- {\sl Resummation of diagrams within DP}. For wetting at a planar wall, it is in fact possible to resum all the diagrams appearing within the DP approximation, eq.\ (\ref{DPW}). This, again, makes use of the idea of effective many-body interactions, and is possible because in a general diagram $\Omega_\mu^\nu$ one can integrate exactly over any final black dot of the zig-zag that is on the wall. Thus, the DP non-local Hamiltonian for wetting in a planar wall, with fixed boundary conditions, is given by
\begin{eqnarray}
\frac{H_{DP}[\ell]}{\Sigma_{\alpha\beta}}=\int\!\!d{\bf s}_\ell\;(1+t\,e^{-\kappa\ell})^2\;\,+\hspace*{3.cm}
\\
\nonumber
\int\!\!\!\!\int\!\!d{\bf s}_1 d{\bf s}_2\;\;\rho({\bf x}_1)\,M({\bf x}_1,{\bf x}_2)\,\rho({\bf x}_2)
\hspace*{1.cm}
\end{eqnarray}
where
\begin{eqnarray}
\rho({\bf x})=e^{-\kappa\ell({\bf x})}+t\,e^{-2\kappa\ell({\bf x})}
\end{eqnarray}
and
\begin{eqnarray}
M({\bf x}_1,{\bf x}_2)=S({\bf x}_1,{\bf x}_2;\bar\ell_{12})+\hspace*{4.5cm}
\\[.2cm] \nonumber
\int\!\!d{\bf s}_3\;M({\bf x}_1,{\bf x}_3)\;S({\bf x}_3,{\bf x}_2;\bar\ell_{32})\;e^{-2\kappa\ell({\bf x}_3)}
\hspace*{1.5cm}
\end{eqnarray}
and $S$ is defined in (\ref{twobody2}). The first integral generates three local contributions: the surface area, $\Omega_1^1$ and $\Omega_2^1$. At leading order, the total two-body interaction $M$ is given by $S$, in which case the $t$ independent terms in the second integral reduce to $\Omega_1^2$. In fact, the two-body term $M$ is very well approximated by
\begin{eqnarray}
M({\bf x}_1,{\bf x}_2)\approx \frac{\;S({\bf x}_1,{\bf x}_2;\bar\ell_{12})\;}{1-e^{-2\kappa\bar\ell_{12}}}
\end{eqnarray}
which shows how the higher-order diagrams resum to give a hard-wall repulsion in the two-body interaction.

6.- {\sl Full diagrammatic structure beyond DP}. We have not attempted to classify diagrams that contain three or more tubes. This is much more cumbersome to do beyond the DP approximation and, with the exception of tricritical wetting studies, is largely of academic interest only. Nevertheless, such structure must exist as can be seen from the following argument. We have shown that the ${\mathcal{X}}$ diagram generates a term $\kappa \ell e^{-2\kappa\ell}$ in the planar binding potential $W_{\pi}(\ell)$ whose coefficient is $\propto t^2$. Naively, this suggests that the mean-field excess free-energy, obtained by minimising $W_{\pi}(\ell)$, contains a higher-order logarithmic singularity $t^4 \ln|t|$. However, such a contribution does not exist as can be seen from the full mean-field calculation for the LGW model. This must mean that, when evaluated at the equilibrium mean-field wetting layer thickness, the term $t^2{\mathcal{X}}$ in $W_{\pi}$ exactly cancels with higher-order diagrams which generate terms of order $t\kappa \ell e^{-3\kappa\ell}$ and $t\kappa \ell e^{-4\kappa\ell}$. Indeed, such diagrams can be readily identified in the perturbation theory. This is strongly suggestive that at least some higher-order diagrams can be grouped together systematically.

Further extensions and applications of this work, including a discussion of correlation function structure and the presence of long-ranged forces, will be discussed in future papers.

AOP is very grateful to Dr.\ A.\ Lazarides and Prof.\ P.M.\ Goldbart for very stimulating conversations. CR acknowledges support from grants MOSAICO (Ministerio de Educaci\'on y Ciencia) and MOSSNOHO (Comunidad de Madrid). NRB acknowledges support from the Portuguese Foundation for Science and Technology (SFRH/BD/16424/2004). JMR-E acknowledges support from the European Commission (MEIF-CT-2003-501042).

\end{document}